# Vehicle State Estimation and Prediction

Xinchen Li, Levent Guvenc, Bilin Aksun-Guvenc

Automated Driving Lab, Department of Mechanical and Aerospace Engineering, Ohio State University

**Abstract**

This paper presents methods for vehicle state estimation and prediction for autonomous driving. A roundabout is chosen to apply the methods and illustrate the results as autonomous vehicles have difficulty in handling roundabouts. State estimation based on the unscented Kalman filter (UKF) is introduced first with application to a roundabout. The microscopic traffic simulator SUMO is used to generate realistic traffic in the roundabout for the simulation experiments. Change point detection based driving behavior prediction using a multi policy approach is then introduced and evaluated for the round intersection example. Finally, these methods are combined for vehicle trajectory estimation based on UKF and policy prediction and demonstrated using the roundabout example.

## 1. Introduction

Connected and autonomous driving [1] including self-driving robo-taxis are becoming more widely available each year. Autonomous driving feedback control loops [2], [3], [4], [5], [6], [7], [8], [9], ;10], [11] and decision-making systems [12], [13], [14], [15] depend on the effectiveness of information collection and learning the knowledge of vehicle motions, including the ego-vehicle and other nearby vehicles. Knowing the information, the autonomous vehicles can estimate the behaviors and future positions of others so as to determine the way of behaving in current traffic scenario. Therefore, the knowledge of vehicles at current moment on motions and states are particularly essential for autonomous driving.

As for autonomous vehicles driving on the road, the sensor suite deployed on them commonly includes GPS, IMU, Lidars, Cameras and Radars. With the information collected from GPS and IMU, the ego vehicle can measure its states, including the global position, the heading angle that shows the orientation, the linear velocity and angular velocity as well as acceleration. With the information collected from those on-board sensors, current states and future state trajectories can be computed and estimated for the ego vehicle. However, the case is a lot different for other vehicles as compared to the ego vehicle. The task for estimating true object states and predict future trajectories of other road users can become a difficult task due to the following reasons. First, detecting sensors such as Lidar, Camera and Radar can provide positioning information and velocity information but unlike ego vehicle with onboard IMU, angular velocity and accelerations are hard to derive simply based on observations from those sensors. Additionally, sensor detections on autonomous vehicle are not exact. The data collection about other road users and objects are noisy which leads to error in sensor detection and perception. This is not truly reliable. And not all of the states of other vehicles are observed or measured. Lastly, the motion of other road users are uncertain, especially in the urban traffic scenario. Unlike highway traffic, the acceleration, velocity and heading angles can change suddenly due to the complex traffic situation. Road users need to be alerted promptly when some unexpected situation possibly takes place in the urban scenario, and this is even more severe when the traffic density is high. Hence, for the autonomous driving planning, it is important to have good estimation and prediction for other road users so that the planning and control of ego vehicle can be more efficient and robust towards the desired control objective.

Vehicle estimation and future state prediction has long been a research problem that people focus on, and a variety of solutions for vehicle estimation and prediction have been proposed. The estimation can be made for vehicles on different types of future behavior and information, especially on the states that are not directly measured, such as angular velocity and lateral velocity. Here we call them the internal states. A lot of methods have been proposed for deriving the internal states from the measured vehicle trajectory. In [16], by assuming the ideal condition where vehicle tires have only pure rolling contact, the authors computed lateral velocity from longitudinal velocity and steering angle based on the single-track model or the so-called bicycle model. In [17], the author compare different motion models for vehicle tracking and apply Kalman filter based method for generating an estimated trajectory to illustrate the accuracy of motion models describing vehicle motions and get rid of the sensor noise. This work provides a good reference for selecting the right motion model to describe vehicle motions under different road conditions. In [18], Real Time Kinematic (RTK) based GPS correlation is applied for vehicle position information collected for estimating vehicle internal states which increase the precession of the collected information. More and more RTK based GPS unit are used for vehicles to improve the accuracy of the positioning system of the vehicle, and this also benefits the vehicle state estimation a lot.

With the fast development of artificial intelligence and neural network method, image based trajectory prediction are also draws a lot of attention. In [19], a graph Long Short Term

Memory (graph-LSTM) based trajectory prediction method is proposed utilizing a series of image and represent the proximity between road agents using a weighted dynamic geometric graph. In [20], a spatio-temporal graph convolution neural network is proposed for predicting human trajectory on road, which can be helpful in preventing pedestrians using road from being hurt by other road users such as vehicle and motorcycles. There are also other Neural Network (NN) based prediction method utilizing LSTM or Recurrent Neural Network (RNN) for the features of processing series of data and predicting future such as [21].

In this chapter, vehicle internal state estimation and vehicle policy prediction methods are introduced for predicting the future state and trajectories for other vehicles in the urban traffic scenarios, especially in the traffic scenario of round intersection. Kalman filter based state estimation and change point detection method based policy prediction are introduced here. Also, a combined method for vehicle trajectory prediction is also proposed in this chapter for better estimating the vehicle trajectory in a future time period, which will assist the decision making of the ego vehicle in the round intersection.

Without loss of generality, we assume that the other road users that the ego vehicle encounter are vehicles. Vulnerable road users such as bicyclist and pedestrians are not considered in this paper. The rest of the paper is organized as follows. State estimation based on the unscented Kalman filter is introduced here with application to a roundabout in Section 2. The microscopic traffic simulator SUMO is used to generate realistic traffic in the roundabout for the simulation experiments of Section 2. Change point detection based driving behavior prediction using a multi policy approach is introduced in Section 3 and evaluated for the round intersection example. Section 4 combines the methods of Sections 2 and 3 for vehicle trajectory estimation based on UKF and policy prediction. The paper ends with the usual conclusions in the last section.

## 2. State Estimation Based on Kalman Filter

With on-board sensors such as GPS and IMU providing information on position and motion, including velocity, acceleration and angular velocity, the future motion and states of the ego vehicle can be predicted. However, for the case where we do not have any information or measurement on the vehicle motion, it is important to have a tool for calculating and estimating the internal states which others do not have access to. Kalman Filter, the famous estimation algorithm, is a powerful and commonly used algorithm for the estimation. The core idea of the Kalman filter and the central operation is the propagation of the Gaussian random variable through the system dynamic model [22]. An original Kalman filter can be used for solving the problem where the system dynamic equations are linear. In reality, however, the vehicle models are nonlinear, so the original Kalman filter is not applicable to solve the trajectory tracking and motion estimation problem. For the nonlinear case, several variations of Kalman filter are proposed. Those include the Extended Kalman filter (EKF) [23] and the Unscented Kalman filter (UKF) [24]. The EKF approximated the states using Gaussian random variables by introducing the first-order linearization of the system dynamic of the nonlinear system. The UKF deals with the non-linearity by representing the state distribution with sampling points, called sigma points that capture true mean and covariance of the Gaussian random variables that describes the states and propagation through the nonlinear function without linearization. Both variations of the Kalman filter work well for the nonlinear case, however, the EKF, as discussed in [24] has some flaws. The first order linearization of the system can introduce large errors for the posterior mean and covariance when the Gaussian random variables are propagated through the linearized state equation. Other than this error caused by linearization, the estimation based on EKF may also has sub-optimal performance, and sometimes will not converge and l will lead to a failure. However, in the UKF, the propagation through dynamic equations of the sample points will not involve additional errors. The posterior mean and covariance will still be captured by the unscented transformation in the UKF. Hence, it out-performs the EKF and is implemented in this work for estimating the vehicle internal states based on the observable measurements and collected data.

## 2.1 The Unscented Kalman Filter

The unscented Kalman filter is a variation of the Kalman filter based on Unscented Transform (UT) that is operated on the sampled sigma points which are used for representing the true mean and covariance of the state representation of Gaussian random variables. Consider a discrete non-linear dynamical system expressed as,

$$x_{k+1} = F(x_k, v_k) \tag{1}$$

$$y_k = H(x_k, n_k) \tag{2}$$

where $x_k$ represents the states of the system at current time. This system could be observable, partially observable or hidden to the external observer and measurement. $v_k$ is the process noise that drives the system or disturbs the state transition of the whole system. $y_k$ is the current measurement which the external observer can access and $n_k$ is the measurement noise caused by the sensors. $F$ is the system dynamic model that is a non-linear function and the measurement function $H$ can also be non-linear or linear in this case.

To realize the Unscented Kalman filter on the non-linear system, the unscented transformation (UT) is introduced. Considering a non-linear system $y = g(x)$ with the n dimension of state vector $x = [x_1, x_2, ..., x_n]^T$ where the initial state of the system is known with its mean and covariance to be $\bar{x}$, $P_x$. A series of $2n+1$ sampling points, called Sigma Points or Sigma Vector and the corresponding weights are generated based on the formulation in Equations (3). The generated Sigma Vector can successfully capture the mean and covariance of the original states.

$$X_0 = \bar{x} \tag{3.a}$$

$$X_i = \bar{x} + (\sqrt{(n+\lambda)P_x})_i \qquad i = 1,2,...,n \tag{3.b}$$

$$X_i = \bar{x} - (\sqrt{(n+\lambda)P_x})_{i-n} \qquad i = n+1,...,2n \tag{3.c}$$

$$w_0^m = \lambda / (n+\lambda) \tag{3.d}$$

$$w_0^c = \lambda / (n+\lambda) + (1-\alpha^2+\beta) \tag{3.e}$$

$$w_i^m = w_i^c = 1/\{2(n+\lambda)\} \qquad i = 1,2,...,2n \tag{3.f}$$

For the UT shown in Equations (3), scaling parameters $\kappa, \alpha$ are introduced. $\alpha$ is the scaling parameter that determines how wide the Sigma Points are spread from the original mean of the states. $\kappa$ is normally set to be 0 or $3-n$. $\beta$ is related to the prior probability distribution of the state variables. As the widely used Gaussian distribution for representing the state variable and noise, $\beta$ is set to be 2 which will provide optimality. Here, the $\lambda = \alpha^2(n+\kappa) - n$ for scaling as well.

Unlike the original KF or EKF which pass through the state mean directly to the dynamic equation, in UKF, the Sigma Points will pass through the dynamic equation so that $2n+1$ Sigma Points representing the prediction of the state are generated.

$$Y_i = g(X_i) \quad i = 0,1,2,...,2n \tag{4}$$

The posterior means and covariances of the prediction Sigma Points are derived with the prediction sigma points along with the weights generated from the UT, as given in Equation (5).

$$\bar{y} \approx \sum_{i=0}^{2n} w_i^m Y_i$$
$$P_y \approx \sum_{i=0}^{2n} w_i^c \{Y_i - \bar{y}\}\{Y_i - \bar{y}\}^T \tag{5}$$

As shown in the above equation, the superscript for weights $w_i$ is related to the mean and covariance. With unscented transformation, the posterior mean and covariance of the states will not be as deviated from the true mean and covariance as they are in EKF due to the error in the 1st order linearization. Hence, it improves the accuracy of state estimation when the dynamic system is non-linear. As provided in [24], the Unscented Kalman Filter based on Unscented Transformation is provided and introduced in Table 1 as Algorithm 1.

| Algorithm 1 Unscented Kalman Filter |
| --- |
| Given state dynamic equation $x_{k+1} = F(x_k, v_k)$ and $y_k = H(x_k, n_k)$.<br><br>Initialize the UKF with:<br><br>$E(x_0) = \bar{x}_0, P_0 = E[(x_0 - \bar{x}_0)(x_0 - \bar{x}_0)]^T, v_k \sim N(0, Q_k), n_k \sim N(0, R)$<br><br>For $k = 1, \ldots$:<br><br>Calculate the Sigma Points of the state:<br><br>$X_{k-1} = [\hat{x}_{k-1}, \hat{x}_{k-1} \pm \sqrt{(n+\lambda)P_{k-1}}]$<br><br>Predictions:<br><br>$X_{k\|k-1} = F(X_{k-1}, v_{k-1})$<br>$\bar{x}_{k\|k-1} = \sum_{i=0}^{2n} w_i^m X_{i,k\|k-1}$<br>$P_{k\|k-1} = \sum_{i=0}^{2n} w_i^c \{X_{i,k\|k-1} - \bar{x}_{k\|k-1}\}\{X_{i,k\|k-1} - \bar{x}_{k\|k-1}\}^T$<br><br>Measurement Update:<br><br>$Y_{k\|k-1} = H[X_{k\|k-1}, n_{k-1}]$<br>$\bar{y}_{k\|k-1} \approx \sum_{i=0}^{2n} w_i^m Y_{i,k\|k-1}$<br>$P_{y,k\|k-1} \approx \sum_{i=0}^{2n} w_i^c \{Y_{i,k\|k-1} - \bar{y}_{k\|k-1}\}\{Y_{i,k\|k-1} - \bar{y}_{k\|k-1}\}^T + R_k$<br>$P_{x_k y_k} \approx \sum_{i=0}^{2n} w_i^c \{X_{i,k\|k-1} - \bar{x}_{k\|k-1}\}\{Y_{i,k\|k-1} - \bar{y}_{k\|k-1}\}^T$<br><br>Estimation:<br><br>$K = P_{x_k y_k} inv(P_{y,k\|k-1})$<br>$\hat{x}_k = \bar{x}_{k\|k-1} + K(y_k - \bar{y}_{k\|k-1})$<br>$P_k = P_{k\|k-1} - KP_{y,k\|k-1}K^T$ |

Table 1. The Unscented Kalman Filter

In algorithm 1, $X_{k-1}$ is the unscented transformed state vector at time k-1. Subscript $k \mid k-1$ represents the derivation from state at step k-1 to k. $X_{k|k-1}$ is the prediction state vector from step k-1 to k, $Y_{k|k-1}$ is the measurement update from step k-1 to k. $\overline{x}_{k|k-1}$ is the true mean of the prediction state vector $X_{k|k-1}$, and $\overline{y}_{k|k-1}$ is the corresponding true mean of $Y_{k|k-1}$. $K$ is the Kalman gain and $\hat{x}$ represents the posterior estimation as the optimal estimated states using the Unscented Kalman Filter.

## 2.2 Motion models for vehicle state tracking and estimation

A good model describing the motion of vehicles is very important for estimating the vehicle states and predicting the future potential trajectories. It is a vital problem of planning and decision making for autonomous vehicle since it is related to the possibility of collision between the ego vehicle and other vehicles.

In order to increase the accuracy and stability of estimation, a motion model is assumed to define and describe the evolution of the vehicle states and dynamic behavior. For simplification, a single-track motion model is commonly used when the kinematic behavior is more focused than the performance of the powertrain. A force analysis model is needed when the powertrain behavior is studied as well as the forces applied on the tire by the ground interaction [50] .Therefore, in different traffic scenarios, different motion models reveal different advantages and disadvantages for tracking  and estimating the vehicle parameters.

In this dissertation, the traffic scenario that draws the most attention is the intersections in an urban environment. Several motion models have been proposed for describing the vehicle dynamic behavior in vehicle turning, for example,  as discussed in [25], the constant turn rate and velocity model (CTRV), constant turn rate and acceleration model (CTRA) as well as the constant curvature and acceleration (CCA) model outperform the constant acceleration model (CA) for describing the vehicle in a curved road situation. Also, the complexity of the model does not make the model a lot better than a relatively simple model. In this work, since true vehicle in urban traffic are not likely to change its speed all the time, and to make sure that the model is simple enough for real time decision making use, the constant turn rate and velocity (CTRV) model is employed and applied for estimating the vehicle states and predicting future vehicle trajectories. For discrete time use, the CTRV model is presented in Equations (6) - (7). The state space for the CTRV model is characterized by the 5-dimensional state tuple given by

$$s_t = [x, y, \theta, v, w]^T \tag{6}$$

In the state vector, $x$, $y$ represent the 2-D planar position information, $x$ for longitudinal position and $y$ for lateral position; $\theta$ is the heading angle of the vehicle; $v, w$ are the linear velocity and yaw angular velocity that captures the vehicle motion state at the current time stamp. Assume the sampling time to be $\Delta t$, the system dynamic Equation then given by

$$s_{t+1} = \begin{bmatrix} x_{t+1} \\ y_{t+1} \\ \theta_{t+1} \\ v_{t+1} \\ w_{t+1} \end{bmatrix} = F(s_t) = \begin{bmatrix} \frac{v_t}{w_t}\sin(w_t\Delta t + \theta_t) - \frac{v_t}{w_t}\sin(\theta_t) + x_t \\ -\frac{v_t}{w_t}\cos(w_t\Delta t + \theta_t) + \frac{v_t}{w_t}\cos(\theta_t) + y_t \\ w_t\Delta t + \theta_t \\ v_t \\ w_t \end{bmatrix}, \quad w \neq 0 \tag{7}$$

When the angular velocity of the vehicle becomes zero, the system dynamical equation becomes

$$s_{t+1} = \begin{bmatrix} x_{t+1} \\ y_{t+1} \\ \theta_{t+1} \\ v_{t+1} \\ w_{t+1} \end{bmatrix} = F(s_t) = \begin{bmatrix} v_t\cos(\theta_t)\Delta t + x_t \\ v_t\sin(\theta_t)\Delta t + y_t \\ \theta_t \\ v_t \\ 0 \end{bmatrix}, w = 0 \tag{8}$$

This model performs well for describing and tracking the vehicle kinematic motion. The vehicle trajectory generated from this model is with the assumption that vehicle is moving on a circular trajectory. The drawback of this type of kinematic model is that the critical dynamics of vehicles can't be revealed while estimating and tracking the motion of the vehicle. But this is sufficient for estimating the hidden states that cannot be observed.

In Equations (6) through (8), no process noise is involved in the system dynamics. When considering process noise that perturbates motion variables, velocity and angular velocity, the system dynamic equation with perturbation and the covariance matrix used for UKF based state estimation are introduced. The noise is mainly noise in acceleration that affects the vehicle velocity and change rate of angular velocity that will disturb the rotational motion of the vehicle. The noise is given as a vector of white noise,

$$\begin{bmatrix} v_a \\ v_w \end{bmatrix}_t \sim N\left(\mu, \mathrm{cov}(v)\right) \tag{9}$$

where $\mu = \begin{bmatrix} 0 \\ 0 \end{bmatrix}$, $\mathrm{cov}(v) = \begin{bmatrix} \sigma_{v_a} & 0 \\ 0 & \sigma_{v_w} \end{bmatrix}$ and it takes place in the process of state dynamics of the motion model which leads to a new state update equation

$$s_{t+1} = F(s_t, v_t) = \begin{bmatrix} \frac{v_t}{w_t}\sin(w_t\Delta t + \theta_t) - \frac{v_t}{w_t}\sin(\theta_t) + x_t + \frac{1}{2}\Delta t^2 \cos(\theta_t)v_a \\ -\frac{v_t}{w_t}\cos(w_t\Delta t + \theta_t) + \frac{v_t}{w_t}\cos(\theta_t) + y_t + \frac{1}{2}\Delta t^2 \sin(\theta_t)v_a \\ w_t\Delta t + \theta_t + \frac{1}{2}\Delta t^2 v_w \\ v_t + v_a\Delta t \\ w_t + v_w\Delta t \end{bmatrix}, \quad w \neq 0 \tag{10}$$

Let G in Equation (11) given below be the noise process function, the covariance matrix of process noise in the UKF state estimation is provided in (12) and will be used for the generation of covariance matrix for predicted state vectors.

$$G = \begin{bmatrix} \frac{1}{2}\Delta t^2 \cos(\theta_t) & 0 \\ \frac{1}{2}\Delta t^2 \sin(\theta_t) & 0 \\ 0 & \frac{1}{2}\Delta t^2 \\ \Delta t & 0 \\ 0 & \Delta t \end{bmatrix} \tag{11}$$

$$Q_k = G\,\text{cov}(v)G^T \tag{12}$$

Using the constant turn rate and velocity (CTRV) motion model with process noise, the UKF based vehicle state trajectory is implemented for estimating the hidden states of other vehicles in the round intersection scenarios, mainly the yaw rate of other vehicles so that the ego vehicle will learn about the change of direction of other vehicles. The same process noise item is also applied when the angular velocity is zero. In the UKF based trajectory estimation, the measurement noise is revealed by directly adding the covariance matrix $R$ to the measurement covariance matrix $P_{y,k|k-1}$. However, the process noise can't be dealt with by directly adding the covariance matrix of process noise to the covariance matrix of the state prediction. Instead, as shown in [55] the state estimation method deals with the process matrix by using the augmented state vector and directly applying the covariance matrix of the noise into the state covariance matrix. In this way, the above mentioned process noise can be taken care of.

## 2.3 Simulation experiment set up and results

The experiment of implementing UKF for vehicle trajectory tracking is tested based on simulation trajectory generated from SUMO environment. The vehicle model for Unscented Kalman Filter based vehicle trajectory tracking is the CTRV model introduced in this chapter. Here, we use one of the experiments for demonstration and show the test results based on SUMO simulation. A vehicle in SUMO travels around the round intersection and generates the path shown in Figure 1.

To simulate the noisy process of the vehicle state transition and the noise generated in the process of measurement, we manually add process noise and measurement noise to the clean trajectory. As is shown in Equation (9), we add a white noise on the velocity and angular velocity, as:

$$\begin{bmatrix} v_a \\ v_w \end{bmatrix}_t \sim N\left(\mu, \mathrm{cov}(v)\right) \tag{13}$$

where $\mu = \begin{bmatrix} 0 \\ 0 \end{bmatrix}$, $\mathrm{cov}(v) = \begin{bmatrix} \sigma_{v_a} & 0 \\ 0 & \sigma_{v_w} \end{bmatrix}$ with $\sigma_{v_a} = \sigma_{v_w} = 0.01$.

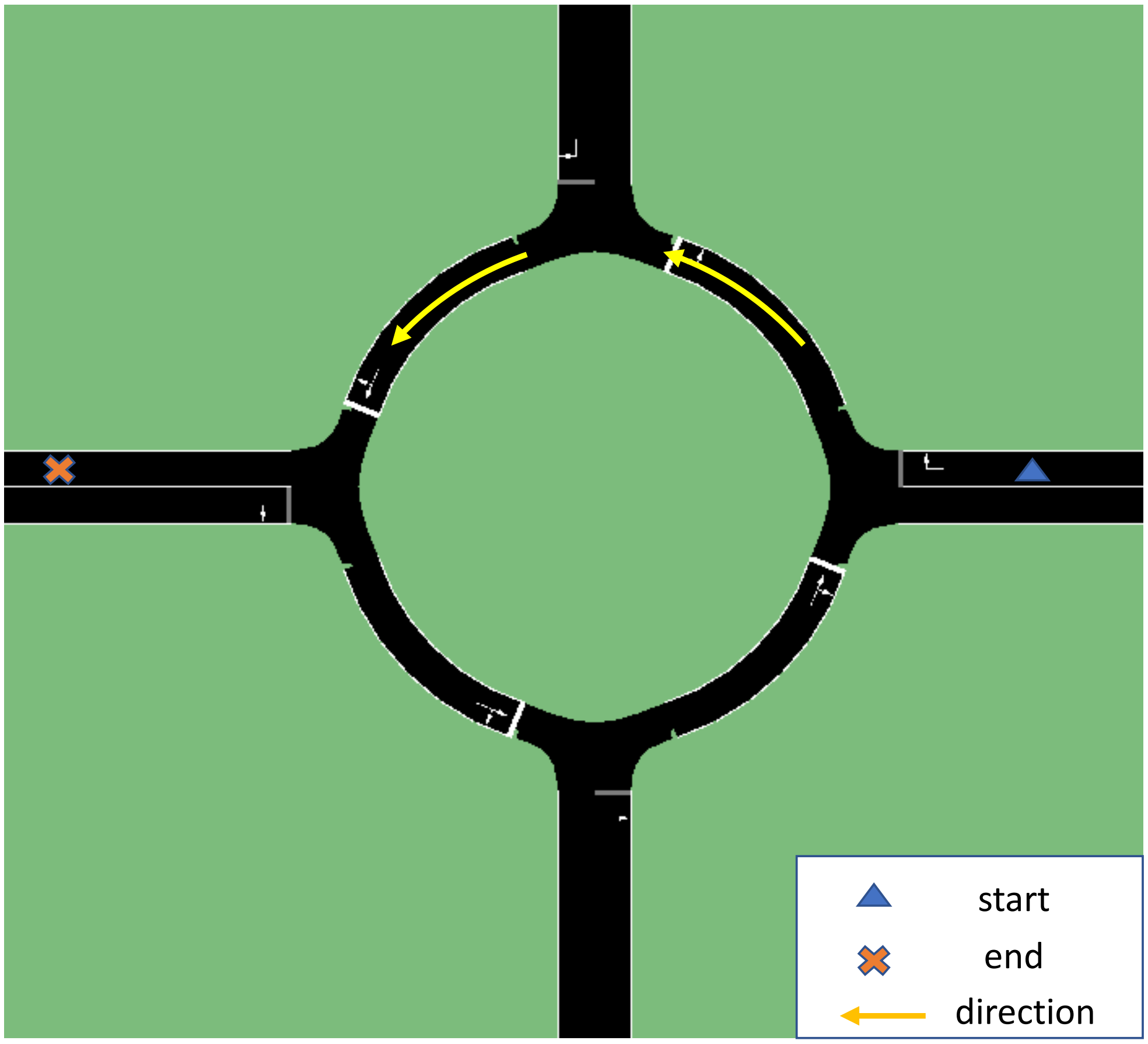

Figure 1 Simulation Environment in SUMO

Each vehicle passes the roundabout intersection, and the clean trajectory is shown in Figure 2.

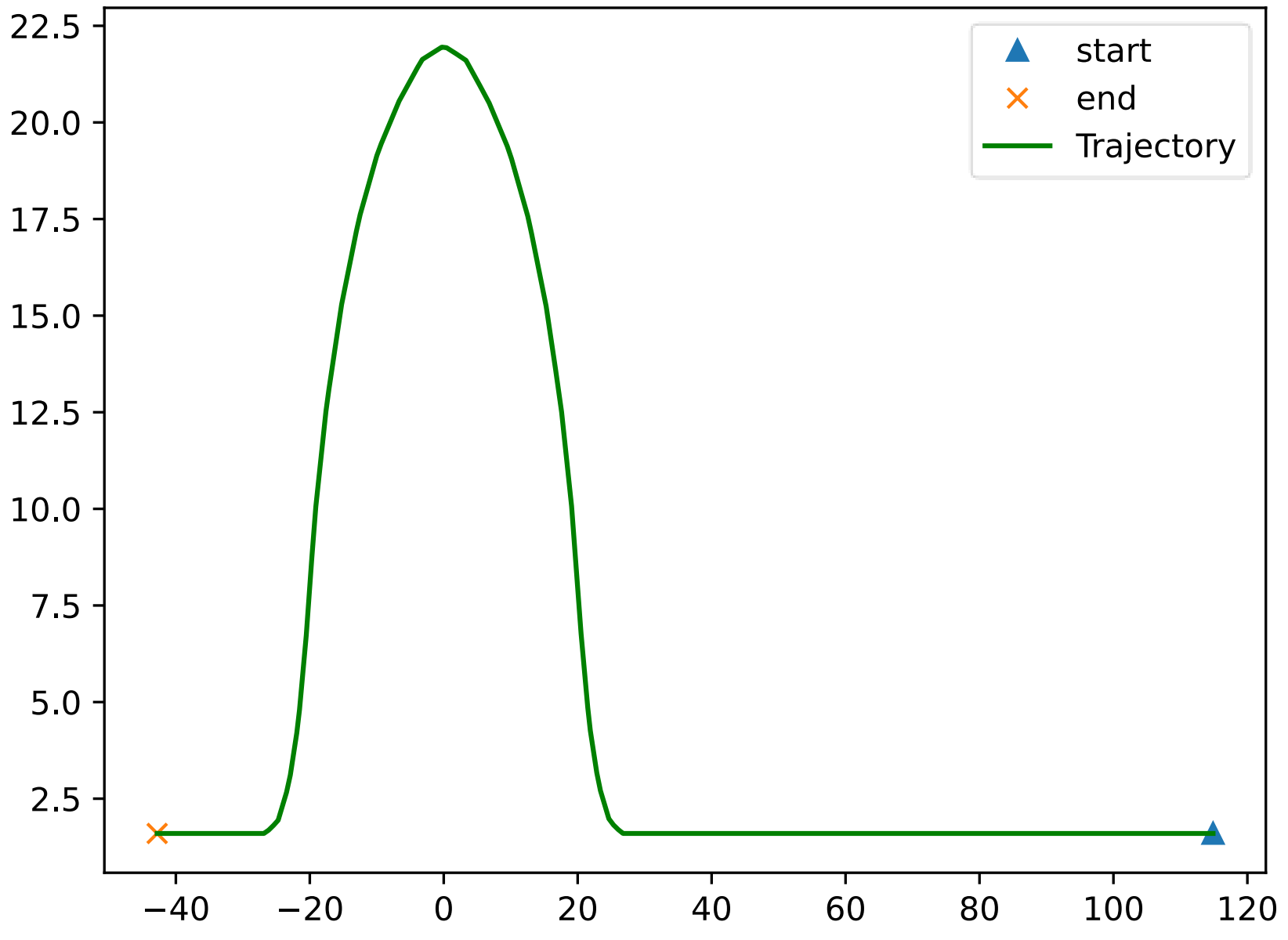

Figure 2. Clean Trajectory of Vehicle Passing the Round Intersection

We also add white noise to the measurement. In the experiment, the measurement is $[x, y, \theta]$, that are vehicle position and the heading angle. The measurement noise is set to be:

$$\begin{bmatrix} n_x \\ n_y \\ n_{theta} \end{bmatrix}_t \sim N\left(\mu_n, \mathrm{cov}(n)\right) \tag{14}$$

where $\mu_n = \begin{bmatrix} 0 \\ 0 \\ 0 \end{bmatrix}$ and $\mathrm{cov}(n) = \begin{bmatrix} \sigma_{n_x} & 0 & 0 \\ 0 & \sigma_{n_y} & 0 \\ 0 & 0 & \sigma_{n_\theta} \end{bmatrix}, \sigma_{n_x} = \sigma_{n_y} = \sigma_{n_\theta} = (0.5)^2$ .

The result of testing is shown in Figure 3, the clean trajectory is recorded from vehicles in SUMO simulation environment shown in green line, and the processed noisy trajectory is shown in blue line while the red line shows the filtered trajectory process by the Unscented Kalman Filter.

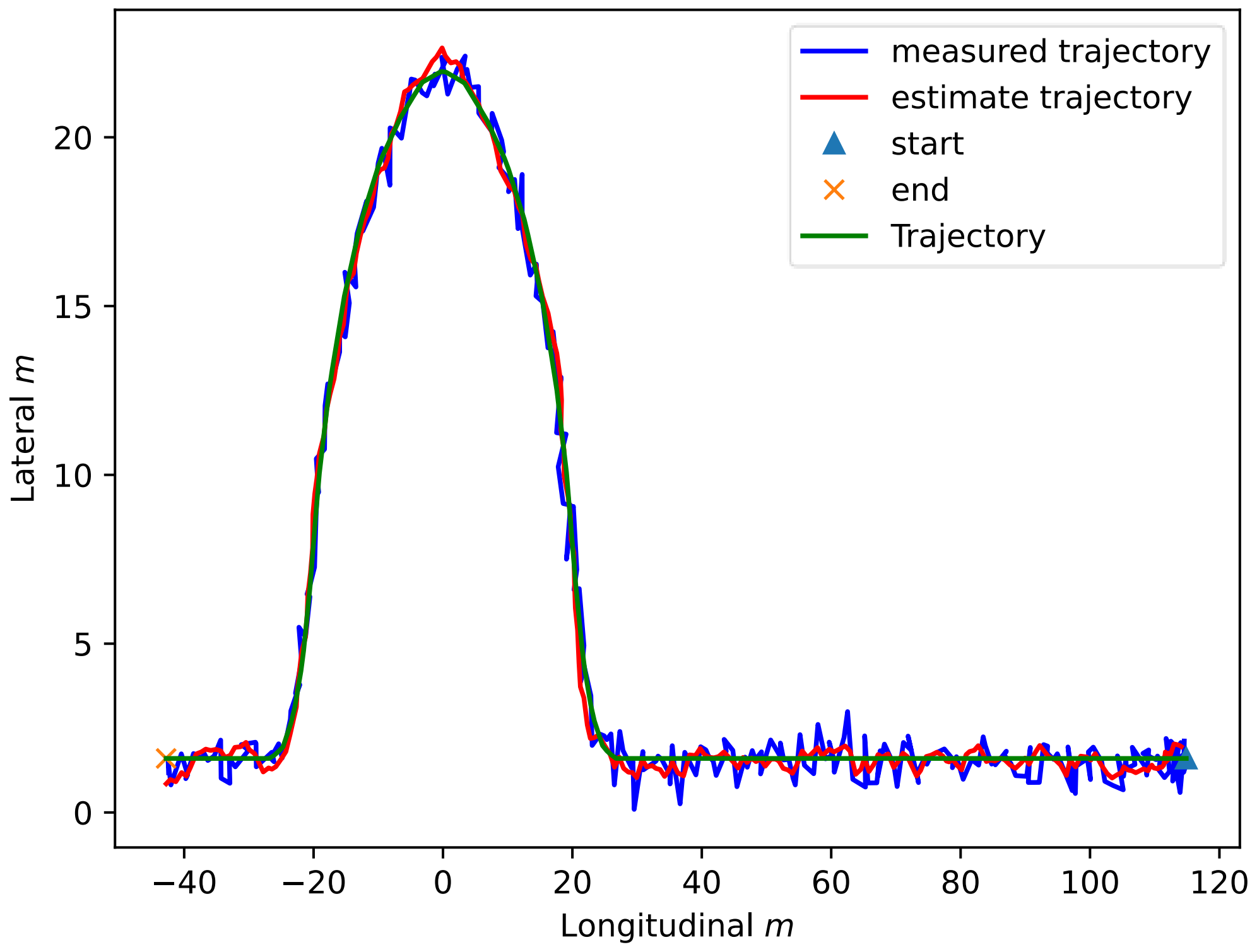

Figure 3. The Result of Vehicle Trajectory Tracking Based on Unscented Kalman Filter

In this trajectory filtering, it is easily seen that the measured trajectory is noisy for the lateral and longitudinal positions and the filtered trajectory can get rid of noise to some extent and converge to the clean trajectory. The filtered trajectory fluctuates the most at the part of the path where the vehicle is turning and it takes longer for the filtered trajectory to converge at the cornering point. For the purpose of estimating hidden states of vehicles, such as velocity and angular velocity, the result is also shown in Figure 4.

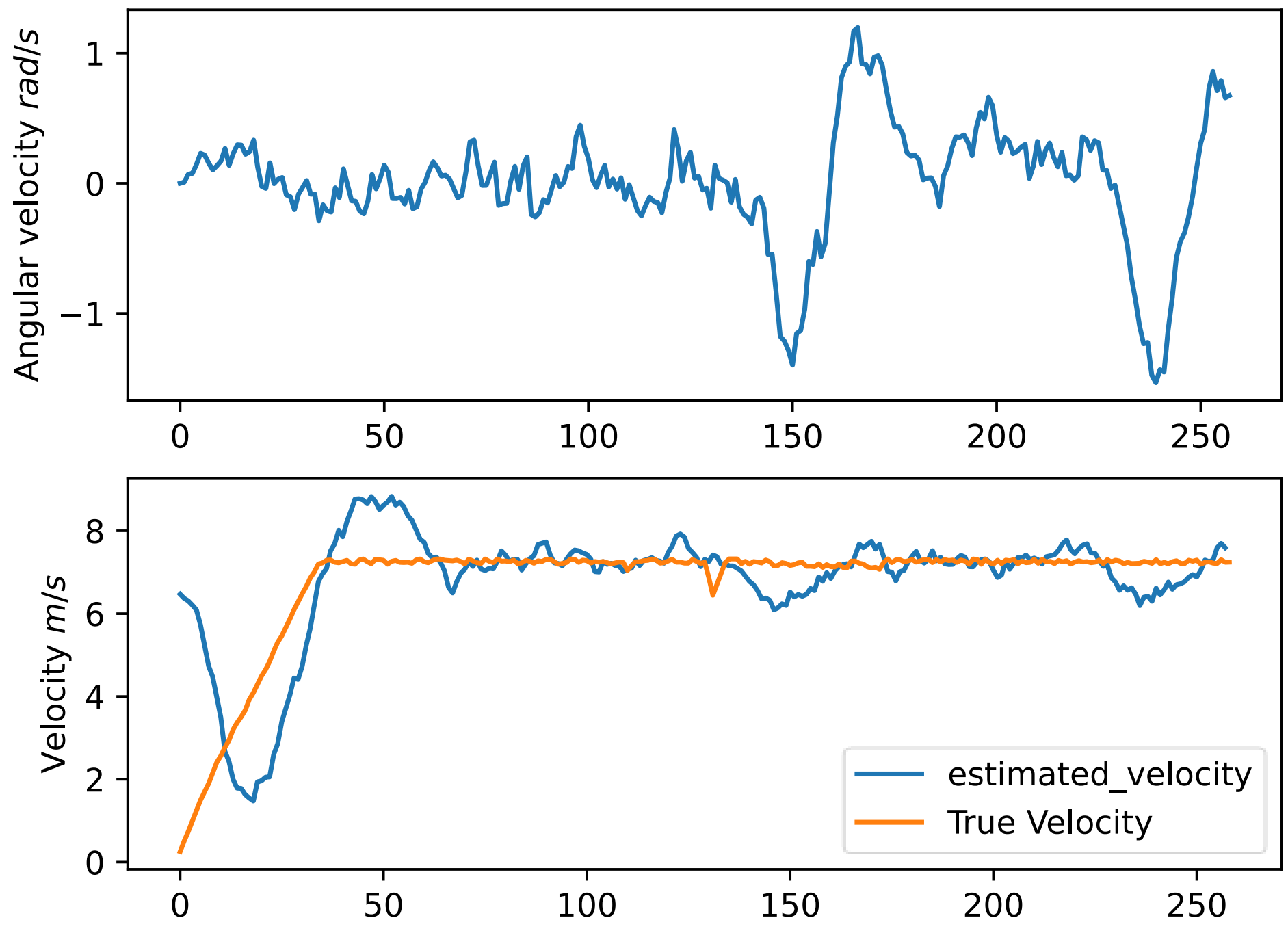

Figure 4 The Estimation of Velocity and Angular Velocity Using Kalman Filter

In the SUMO environment, the vehicle's speed is accessible, hence we use it for comparison to illustrate the performance of the UKF based vehicle trajectory estimation. The initial condition set for estimation is the normal drive speed of the vehicle. Hence it starts with large errors, however, it converges to the true velocity soon and tracks the true velocity. The estimation of the angular velocity is also presented in Figure 4. We do not have access to the angular velocities of other vehicles, hence the UKF based state estimation provides the tool for estimating the hidden states such as angular velocity. This test has been conducted for 3 vehicles' trajectories in the simulation and the test results are listed in

| | Average lateral error\m | Max lateral error\m | Average longitudinal error\m | Max longitudinal error\m | Average Euclidean distance\m | Max Euclidean distance\m |
|---|---|---|---|---|---|---|
| $V_1$ | 0.27 | 0.97 | 0.27 | 1.41 | 0.43 | 1.43 |
| $V_2$ | 0.22 | 1.24 | 0.21 | 0.68 | 0.35 | 1.25 |
| $V_3$ | 0.26 | 0.94 | 0.29 | 1.55 | 0.43 | 1.63 |

Table 2. The Result of Tests on Three Vehicles in The Simulation

## 3. Change Point Detection Based Behavior Prediction

In this section, a policy prediction method leveraging Bayesian change point detection technique is introduced. This method was first introduced in [26] for detecting vehicle in highways to determine the current behavior of the observed vehicles and to detect the anomaly behavior performed by the observed vehicles. In this dissertation, this method was modified by introducing the Kalman Filter based state estimation for generating a clean vehicle trajectory history for assisting of the behavior prediction and a new method for solving the computation of likelihood value for different vehicle policy under the scenario of the round intersection. Also, proposed as a major contribution of this chapter, our modified method predicts the future dynamic behavior of the observed vehicle using the vehicle hidden states estimated by the Kalman Filter with the CTRV model. The test results based on vehicles simulated in SUMO environment are also presented in this section as a minor contribution.

In the previous work of [26], the authors presented an integral method for decision making with the behavior detection. In this paper, we just utilize the method for behavior prediction and combine it with vehicle motion model for predicting vehicle behavior in the near future so that the estimated trajectory derived from the prediction method can assist a decision making algorithm.

### 3.1 Problem formulation for vehicle behavior prediction

In the urban traffic scenario, vehicles around the ego vehicle will behave differently in order to respond to current traffic situation, which makes it a complex task for the ego autonomous vehicle to make decisions in order to realize efficient and safe driving. Hence, it is very helpful to make good prediction of other vehicles' trajectories and behaviors within the observation range of the ego vehicle.

Let $V$ denote the set of vehicles near the ego-vehicle and in the range of on-board sensors of the autonomous ego vehicle. Also include the ego-vehicle into the set, $v_e$ represents the ego vehicle and $v_i, i \in \{1,...,n\}$ represents the $n$ surrounding vehicles such that all the vehicles considered are in the set $V$, given by

$$V = \{v_e, v_1, v_2, ..., v_n\} \tag{15}$$

Each $v_i \in V$ is described by a 5-tuple of state vectors $s_i = (x, y, \theta, v, w)_i$ that represents planar position of the vehicle on the road, heading of the vehicle and also motional variables linear velocity $v$ and angular velocity $w$. At each time t, vehicle $v_i$ will select an action $a_t^{v_i} \in A$ from an action set to drive the vehicle and update its state vector from $s_t^i \to s_{t+1}^i$, both of the states are also included in the set of states $S$.

The vehicle motion model is represented by a conditional probability function *T*,

$$T(s_t, a_t, s_{t+1}) = p(s_{t+1} \mid s_t, a_t) \tag{16}$$

Note that in the context, $s_t \in S$ includes all the state of the vehicles in the vehicle set $V$, and similarly, $a_t \in A$ denotes all the actions that vehicles might take at time *t*. Observations with uncertainty on the vehicles in $V$ is also modeled as

$$Z(s_t, o_t) = P(o_t \mid s_t) \tag{17}$$

$o_t$ as the observation, is also the set of observations that the ego vehicle can acquire through the on-board sensors. Each of the observations made over the individual vehicles $v_i$ is $o_t^i$. Due to the diversity of sensor systems, the observations of other vehicles are also various and need to be processed. In this dissertation, measurements and observations are not the critical part, hence, the observation and the processing of the observation will not be covered here. As was presented in the previous section, the observation used for this part is just partial vehicle states information and hidden variable such as angular velocity not accessible from the ego vehicle. Drivers of the vehicles tend to drive in the manner that maximize some driving demands that can lead to the goal destination or speed in the shortest time. There are also uncertainties contained in the driving behaviors. The driving model is represented as

$$D(s_t, o_t^i, a_t^i) = p(a_t^i \mid s_t, o_t^i) \tag{18}$$

The model in (18) shows the driver behavior is based on current states of all the vehicles and the observations made. Considering the uncertainties of the vehicle states, the vehicle states at time t is modeled and represented by probability distribution $P(s_t^i)$, for $i = 1, ..., n$. The overall states for all the vehicles in $V$ are shown as $P(s_t)$. While vehicles are driving on the road, the states keep transiting based on the transition function $T$, the observation function $Z$ and the driver model $D$ as

$$P(s_{t+1}) = \int_X \int_O \int_A P(s_{t+1} \mid s_t, a_t) P(a_t \mid s_t, o_t) P(o_t \mid s_t) P(s_t) da_t do_t ds_t \tag{19}$$

The sets $S$ refers to all the states of the vehicles in the vehicle set, $O$ represents the set of observations and $A$ contains all the possible actions that vehicle can take under traffic scenarios. With the assumption that each of the vehicles in the set are independent from others, (19) can be transformed as

$$P(s_{t+1}) = \prod_{v_i \in V} p^i(s_{t+1}^i, s_t^i, a_t^i, o_t^i) da_t^i do_t^i ds_t^i \tag{20}$$

$$p^i(s_{t+1}^i, s_t^i, a_t^i, o_t^i) = P(s_{t+1}^i \mid s_t^i, a_t^i) P(a_t^i \mid s_t^i, o_t^i) P(o_t^i \mid s_t^i) P(s_t^i) \tag{21}$$

To predict the future state distribution $P(s_{t+1})$, the prediction of driver model $D(x_t, o_t^i, a_t^i) = p(a_t^i \mid x_t, o_t^i)$ is the main task to solve in this section.

### 3.2 The Multipolicy Approach for driver model prediction

As discussed in [26], commonly seen way of traffic participants will behave in a regular and predictable manner and do not seem to change in a quick, sudden manner such that other vehicles cannot respond. The vehicle motions cannot be changed without any preceding stages either. It would be a good idea to leverage the vehicle state trajectory in a period of timestamps and reason the possible vehicle behavior out of the observed vehicle trajectory. Then, one should find the *maximum a posteriori* (MAP) estimated policy based on likelihood estimation.

The prediction is based on latent vehicle policies that the vehicle on road can possibly take for driving. A vehicle is trying to achieve the goal of driving. This goal can be quantified as a real-time reward function,

$$r : S \times A \to R \tag{22}$$

The reward function depends on the demand of the vehicle. Based on vehicle drivers demand on the road, it may be related to safety issue, velocity, comfort and time to destinations and other concerns that vehicles and their drivers have on the road. A policy is a mapping from current state of the vehicle, current observation to the available action set.

$$\pi : S \times Z \to A \tag{23}$$

An optimal policy in this case, is the policy that maximizes the expected sum of real-time reward over some time horizon, H, as shown in the Equation (24).

$$\pi^* = \arg\max_{\pi} E[\sum_{t=t_0}^{H} R(s_{t_0}, \pi(s_t, o_t))] \tag{24}$$

For decision making problem, the goal of finding the optimal policy is to make sure that the vehicle can drive safely and efficiently. In this section, the goal, instead, is to estimate the most likely policy that the observed vehicle is taking.

To make this estimation problem tractable, as discussed in the previous part, a set of latent discrete policy will be introduced first. The policy set contains the policy that covers hand-engineered policies specifically designed for the target traffic scenario. Here, the policies in the set are high level vehicle control policies that give instructions to the vehicle on what kind of behavior they should perform. An explicit low-level control should be applied for policy execution. This allows a broad range of control algorithms to be implemented and is not in the scope of this work.

Assuming that each vehicle $v^i \in V$, at any time of driving under the current traffic scenario, is executing some policy from the pre-defined policy set, expressed as,

$$\pi^i \in \prod \text{ for } i = 1, 2, ..., n \tag{25}$$

where $\pi^i$ is the policy that vehicle $v^i$ is currently executing and $n$ is the number of vehicles that are not the ego vehicle in the set. In general, the policy itself can be parametrized as a function with respect to current states of the whole set of vehicles $s_t$, including the observed states and hidden states within the state vectors, as well as parameter vector $\theta_i$. This parameter vector can be used for capturing the features of driving, for example, the aggressiveness of the driver. The dynamic limit of the current observed vehicle may lead to different type of vehicle state trajectories however under the same policy from the set. Since the action the driver selects is generated from the policy, the driver model of vehicle $v^i$ in Equation (18) is now expressed as

$$D(s_t, o_t^i, a_t^i, \pi_t^i) = P(a_t^i \mid s_t, o_t^i, \pi_t^i) P(\pi_t^i \mid s_t, o_{1:t}) \tag{26}$$

The first part shows that the action the driver might take is depending on the current state of all the vehicles, considering both other vehicles and the ego vehicle, and the observations and policy. The policy is determined by the current state with the series of observation $o_{1:t}$. The derivation of the conditional probability $P(\pi_t^i \mid s_t, o_{1:t})$ is the core problem to be solved in this section, and its solution will be presented later in this section. Updating the driver model based on policy, the state probability distribution evolution equation shown in Equation (21) can now be approximated by

$$\begin{aligned} p^i(s_{t+1}^i, s_t^i, a_t^i, o_t^i, \pi_t^i) = P(s_t^i) P(o_t^i \mid s_t^i) P(s_{t+1}^i \mid s_t^i, a_t^i) \\ P(a_t^i \mid s_t^i, o_t^i, \pi_t^i) P(\pi_t^i \mid s_t, o_{1:t}) \end{aligned} \tag{27}$$

The overall state probability distribution is then a joint-probability distribution between the state distribution of the ego vehicle and other vehicles in the set, given by

$$P(s_{t+1}) = P(s_{t+1}^e, s_{t+1}^{v_i}), i = 1, 2, ..., n \text{ and } v_i \neq e \tag{28}$$

$$\begin{aligned} P(s_{t+1}) = \int_{S^e} \int_{O^e} P^e(s_{t+1}^e, s_t^e, a_t^e, o_t^e, \pi_t^e) ds_t^e do_t^e \\ \prod_{v^i \in V, v^i \neq e} \left[ \sum_{\Pi} \int_{S^v} \int_{O^v} P^{v_i}(s_{t+1}^{v_i}, s_t^{v_i}, a_t^{v_i}, o_t^{v_i}, \pi_t^{v_i}) ds_t^{v_i} do_t^{v_i} \right] \end{aligned} \tag{29}$$

The ego-vehicle is assumed to be in full control. Therefore, the state distribution update can be derived based on the policies determined by the decision making algorithm implemented for the ego-vehicle.

The future states of other vehicles depend on the policies which we do not know. Thus, the policy shall be predicted for the conditional probability

$$P(\pi_t^i \mid x_t, o_{1:t}) \tag{30}$$

Then, a forward simulation with corresponding low-level controller and appropriate motion model shall be carried out to obtain the future trajectories based on the current estimated policy. The trajectory prediction method used here utilize a Bayesian Change Point detection method leveraging the measured trajectories of the target vehicle. To segment a tracked vehicles state trajectory over a past period of time, the change point detection method proposed in [27], Change Point Detection using Approximate Model Parameters (CHAMP) algorithm, is adopted since it takes advantages of the feature of change point detection and applies it to segment data that are not only generated from random processes, but data obtained by observing vehicles motion based on different types of motion. Since the vehicle motion is continuous without sudden jump if the robot is executing some policy in a segment of the trajectory, the change of policies can be detected given a set of available policies $\prod$ and a series of observed data of given vehicle, denoted as $o_{1:n} = \{o_1, o_2, ..., o_n\}$.

The CHAMP infers the *maximum a posteriori* (MAP) set of times, $\tau_1, \tau_2, ..., \tau_m$, at which change points between policies have occurred, and this will segment the time series into $m+1$ data segments corresponding to policies executed on those data segments. The $i^{th}$ observation segment $o_{\tau_i+1} : o_{\tau_{i+1}}$ is detected associated with a certain policy from the pre-defined policy set. In the following part, the CHAMP algorithm is introduced and illustrated as to how it works for detecting the change point between different data associated with corresponding policies.

The change point positions are also described by probability distribution and can be viewed as a Markov chain where the transition probabilities between neighboring change points are a function of time,

$$P(\tau_{i+1} = t \mid \tau_i = s) = g(t-s) \tag{31}$$

The probability distribution function (pdf) $g(\bullet)$ is a function over the length of data segment, and we denote the cumulative probability distribution (cdf) of the function $g(\bullet)$ as $G(\bullet)$. In the algorithm of CHAMP, the truncated Gaussian function over the length of data segment is used as:

$$g(t) = \frac{\frac{1}{\sigma}\phi(\frac{t-\mu}{\sigma})}{1-\Phi(\frac{\alpha-\mu}{\sigma})} \tag{32}$$

$$G(t) = \Phi(\frac{t-\mu}{\sigma}) - \Phi(\frac{\alpha-\mu}{\sigma}) \tag{33}$$

where $\phi$ is the standard normal distribution pdf, $\Phi$ is its cdf, and the $\alpha$ is a parameter that should be defined based on different cases for using this change point detection algorithm that defines the minimum segment length for the change point detection [27].

Another important factor for segmenting data based on different polices associated with the data segment is the confidence that the series of data is generating under the policy $\pi$, defined as policy evidence:

$$L(s,t,\pi) = P(o_{s+1:t} \mid \pi) = \int P(o_{s+1:t} \mid \pi, \theta) p(\theta) d\theta \tag{34}$$

In the above equation, the policy confidence of policy $\pi$ over the data segment starting from time $s$ to time $t$ is shown. For the purpose of efficient computation and avoiding marginalizing over the parameters in the conditional probability of policy, CHAMP makes an approximation by using logarithm of the policy evidence in Equation (34) as the policy evidence for the data segment via the Bayesian Information Criterion as:

$$\log L(s,t,\pi) \approx \log P(o_{s+1:t} \mid \pi, \hat{\theta}) - \frac{1}{2} k_{\pi} \log(t-s) \tag{35}$$

where $k_{\pi}$ is the number of parameters for the policy $\pi$ depending on the type of pdf describing the policy function; $\hat{\theta}$ is the estimated parameters for policy $\pi$. This BIC approximation for policy evidence does not require marginalizing of the probability over the parameter and it penalizes the complexity of the model with the part $-\frac{1}{2} k_{\pi} \log(t-s)$, with respect to the parameters of the model.

Another advantages of using this BIC for computing the policy confidence is making full use of the series nature of data since the BIC equation of (2-35) is only applicable when the size of data is way larger than the number of parameters. The computation of BIC only requires fitting different policies into the observed data segments and getting the maximum likelihood estimation (MLE).

According to [28], the distribution $C_t$ over the most recent change point before time *t* can be estimated is by re-gaining the Viterbi path by applying an online Viterbi path algorithm and Bayesian filtering recursively. A Viterbi path is the path of the maximum posteriori probability estimate of the most likely sequence of hidden states [29]. In our case, the hidden state is the change point detected. Here are the terms and definitions used for finding the MAP of the most recent change point before time *t*.

$$P_t(j,\pi) = P(C_t = j, \pi, \varepsilon_j, o_{1:t}) \tag{36}$$

This probability distribution shown in Equation (2-36) is the probability of the last change point that occurs at time *j* before *t* associated with policy $\pi$

$$P_t^{MAP} = P(\text{CP at } t, \varepsilon_t, o_{1:t}) \tag{37}$$

and shows the maximum a posteriori choice of a change point occurring at time $t$. Equations (36), (37) result in the following equations:

$$P_t(j,\pi) = (1 - G(t - j - 1))L(j,t,\pi)P(\pi)P_j^{MAP} \tag{38}$$

$$P_t^{MAP} = \max_{j,\pi}\left[\frac{g(t-j)}{1 - G(t - j - 1)} P_t(j,\pi)\right] \tag{39}$$

The $\varepsilon_j$ in the above equations represents an event which the change point based on MAP choice occurs prior to time *j* given the condition that the change point however is at time *j*. At any time *t* in the time series, the Viterbi path can be recovered by maximizing the MAP choice $P_t^{MAP}$ with respect to the pair $(j, \pi)$ as $j$ is the one step previous change point for the data. The whole process will keep repeating until the first data point of the whole data series is reached. The whole Viterbi path consists of all the change points that are derived from the optimization process. The change point will be used as index for segmenting the data into segments.

### 3.3. Behavior prediction for round intersection

In this part, the task of predicting vehicle policy in the round intersection is shown and the solution for estimation is shown for round intersection which is the traffic scenario of interest in this work. Similar to the method proposed in [26], the way of fitting policy to a specific data segment is based on the likelihood value of different policies.

The policy evidence or the policy confidence, as is introduced in previous context, is the core item for determining the associated policy of a given data segment. Note that a series of vehicle state data observed from a vehicle $v$ described the partial state trajectory as $o_{1:t} = \{o_1, o_2, ..., o_t\}$. Given that the trajectory has been detected and $m$ change points has been detected, the last data segment in the data series is $o_{m+1:t}$. Based on Equation (34) and (35), the policy likelihood for the given data segment can be computed with:

$$L(\pi) = \max_{\pi \in \prod, \hat{\theta}} P(o_{m+1:t} \mid \pi, \hat{\theta}) \tag{40}$$

$$\log L(\pi) = \max_{\pi \in \prod, \hat{\theta}} \log P(o_{m+1:t} \mid \pi, \hat{\theta}) \tag{41}$$

The posterior probability of the maximum likelihood estimation is derived with the observed data series and the forward simulation vehicle trajectory under certain policies.

The forward simulation of a vehicle generates a future potential trajectory given $\pi \in \prod$, and it takes advantages of the typical geometry structure of a round intersection. A round intersection can be decomposed into three main parts, the straight lanes connecting to the roundabout, the part of road where the straight lanes are connecting to the round intersection and the round intersection, vehicles will behave differently during the drive on those three types of road segments. To be clear, the vehicle policies during the in-lane driving or high-way like driving are not considered in this work and vehicle behaviors such as lane changing are not the goal for prediction. Mainly two kinds of policies are to be predicted: when the vehicle is driving in lane no matter on the straight road or in the round intersection, the vehicle behavior is defined to be under the policy of *lane keeping*. When the vehicle is trying to enter a round intersection or trying to leave the round intersection into a straight connected road, the policy is defined to be *merge*, which means the vehicle is getting in contact with the connection point between the round intersection and the straight road and approach to a different type of road. Hence, the whole policy set with the driving style as the parameter is given by:

$$\{lane\ keep \cup\ merge\} \tag{42}$$

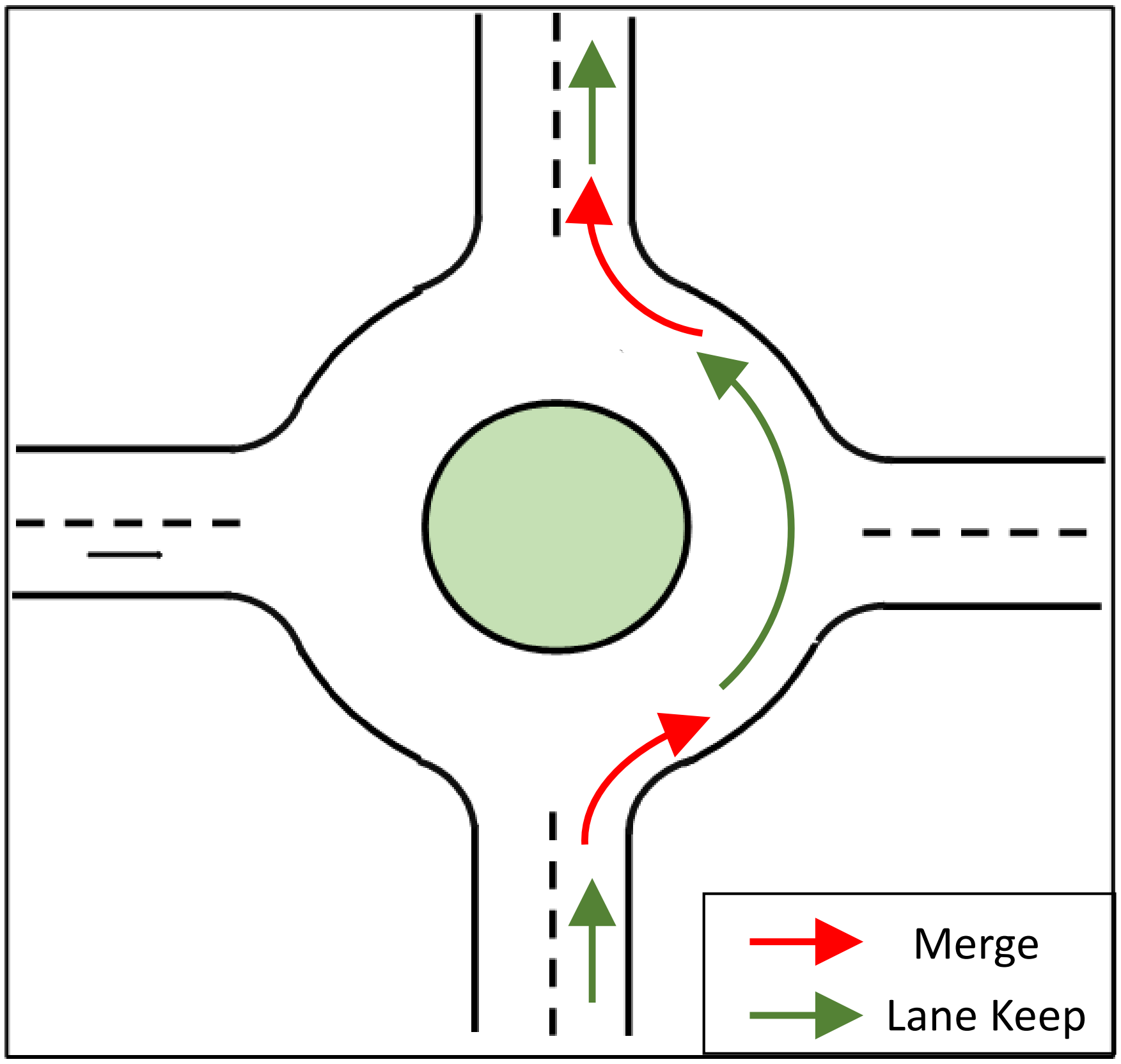

Figure 5 A Demonstration of Policies When a Vehicle is Driving Through a Round Intersection

As demonstrated in the previous context and Figure 5, different policies in the policy set are shown in the figure. A vehicle is passing a round intersection starting from the bottom of the figure to top of the figure along the road. The vehicle first drives in the straight road segment, then merges into the round intersection. After entering the round intersection, it follows the single lane in it until it reaches the target exiting area to exit it and merge back to the straight road segment. The whole process can be divided into five sub-segments as shown in Figure 5, and can be described by the policies in (42) as

$$\begin{aligned} &\{lane\ keep\} \rightarrow \{merge\} \rightarrow \{lane\ keep\} \\ &\rightarrow \{merge\} \rightarrow \{lanekeep\} \end{aligned} \tag{43}$$

To be easily noted, the segments with policy *lane keep* are with the color of green and those with policy *merge* are colored by red.

And the likelihood computation, as shown in Equations (40) and Equation (41), is approximated by finding the likelihood value based on normal distribution with the observed trajectories as parameters:

$$P(o_{m+1:t} \mid \pi, \theta) = N(o_{m+1:t}; T^{\pi,\theta}, \sigma I) \tag{44}$$

Yet, the observed trajectory is a series of state vectors. Each state vector, based on the observation model, should contain vehicle pose information. In this work, the observation of other vehicles is $O_t^v = (x_v, y_v, \theta_v)$, which are longitudinal position $x_v$, lateral position $y_v$ as well as heading angles of the vehicle $\theta_v$. Hence, fitting the whole series of observed state trajectory into the normal distribution will lead to a matrix gaussian distribution and requires computation for high dimensional data. To make the computation more efficient, we made an approximation such that the correlation between the three state variables are ignored and the whole state trajectory is split into three trajectories containing $x_v, y_v$ and $\theta_v$, denoted as $T_{x_v}^{\pi,\theta}, T_{x_y}^{\pi,\theta}, T_{\theta_v}^{\pi,\theta}$. Therefore, Equation (41) is represented as:

$$\begin{aligned} \log L(\pi) &= \max_{\pi\in\Pi,\hat{\theta}} \log P(o_{m+1:t} \mid \pi, \hat{\theta}) \\ &\approx \max_{\pi\in\Pi,\hat{\theta}} \left\{ \log P(x_v,_{m+1:t} \mid \pi, \hat{\theta}) + \log P(y_{v,m+1:t} \mid \pi, \hat{\theta}) + \log P(\theta_{v,m+1:t} \mid \pi, \hat{\theta}) \right\} \end{aligned} \tag{45}$$

$$\widehat{\pi_v}^* = \arg\max_{\pi\in\Pi} \log L(\pi) \tag{46}$$

By computing the log likelihood value as the BIC, the segments will be determined with the policy most likely $\widehat{\pi_v}^*$ from the policy set.

### 3.4 Simulation and test results of policy prediction

In this test, the policy prediction method is used to determine the vehicle's policy in different trajectory segments given a series of vehicle trajectory data. In this work, the tests have been done on clean trajectory directly collected from SUMO environment and the noisy trajectory obtained by adding measurement noise to the clean trajectory. The results discussed below. First, the results on policy prediction based on change point detection are presented. Same as the previous section, the tests are on recorded clean vehicle trajectory passing a roundabout intersection. The vehicles, based on the trajectory data, are determined for executing a policy from the policy set. For the test, the set of parameters for the change point detection based policy prediction is also required. For the segment length distribution in Equations (32) and (33), the mean length of the segment is set to be 50, and the minimum segment length is set to be 25. A result of prediction is shown in Figure 6 and Figure 7. The vehicle travels following the direction of arrows. At the cornering part where vehicle is entering or exiting the round intersection area, the policy prediction algorithm determines that the policy will be *merge.*

While in the other parts of the trajectory, vehicle policy is determined to be lane follow as we have illustrated in previous section that both in the straight lane segment and in the lane of roundabout intersection area, the vehicle policy is set to be lane follow.

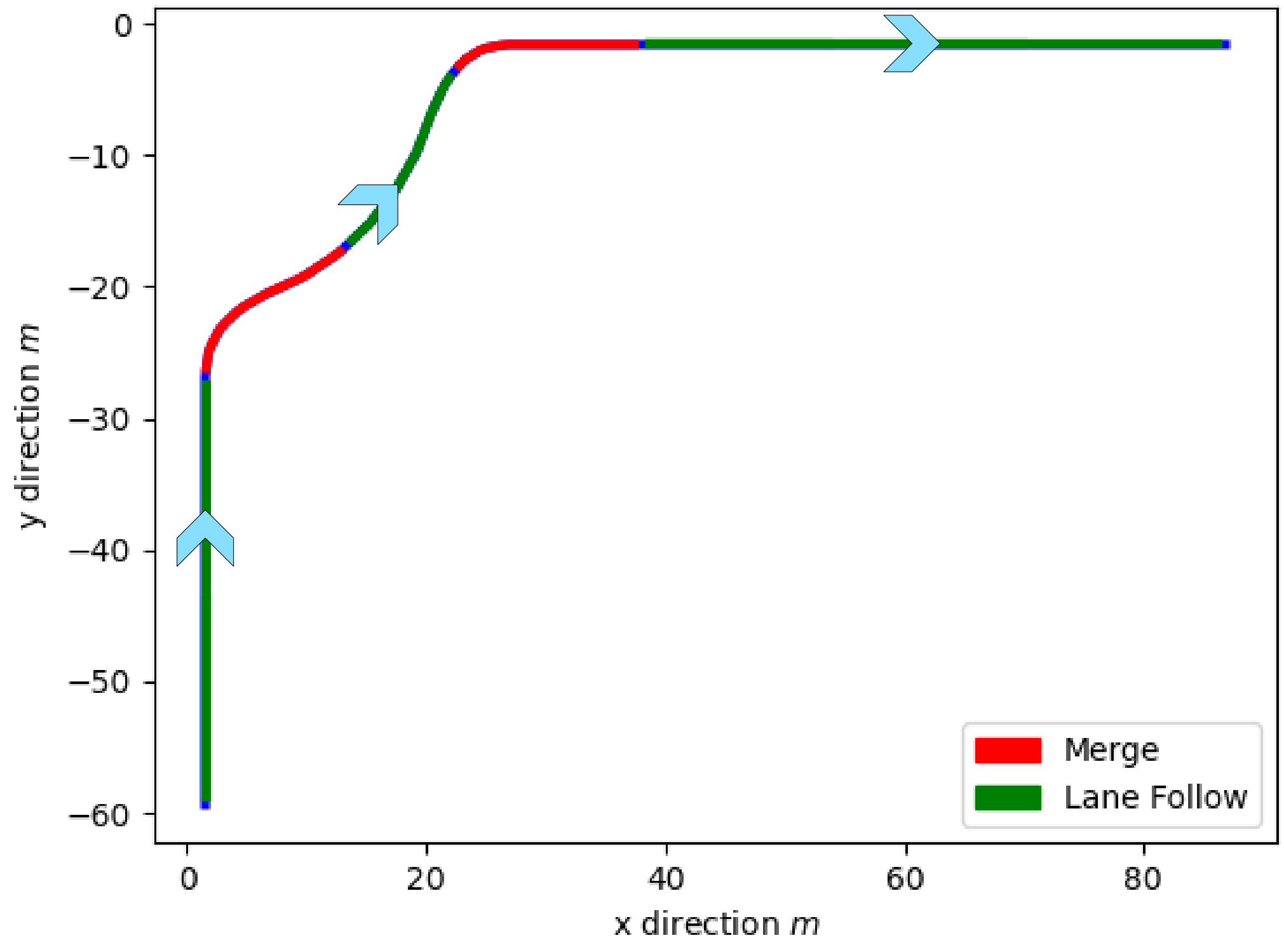

Figure 6 The Policy Prediction Results on a Vehicle Trajectory

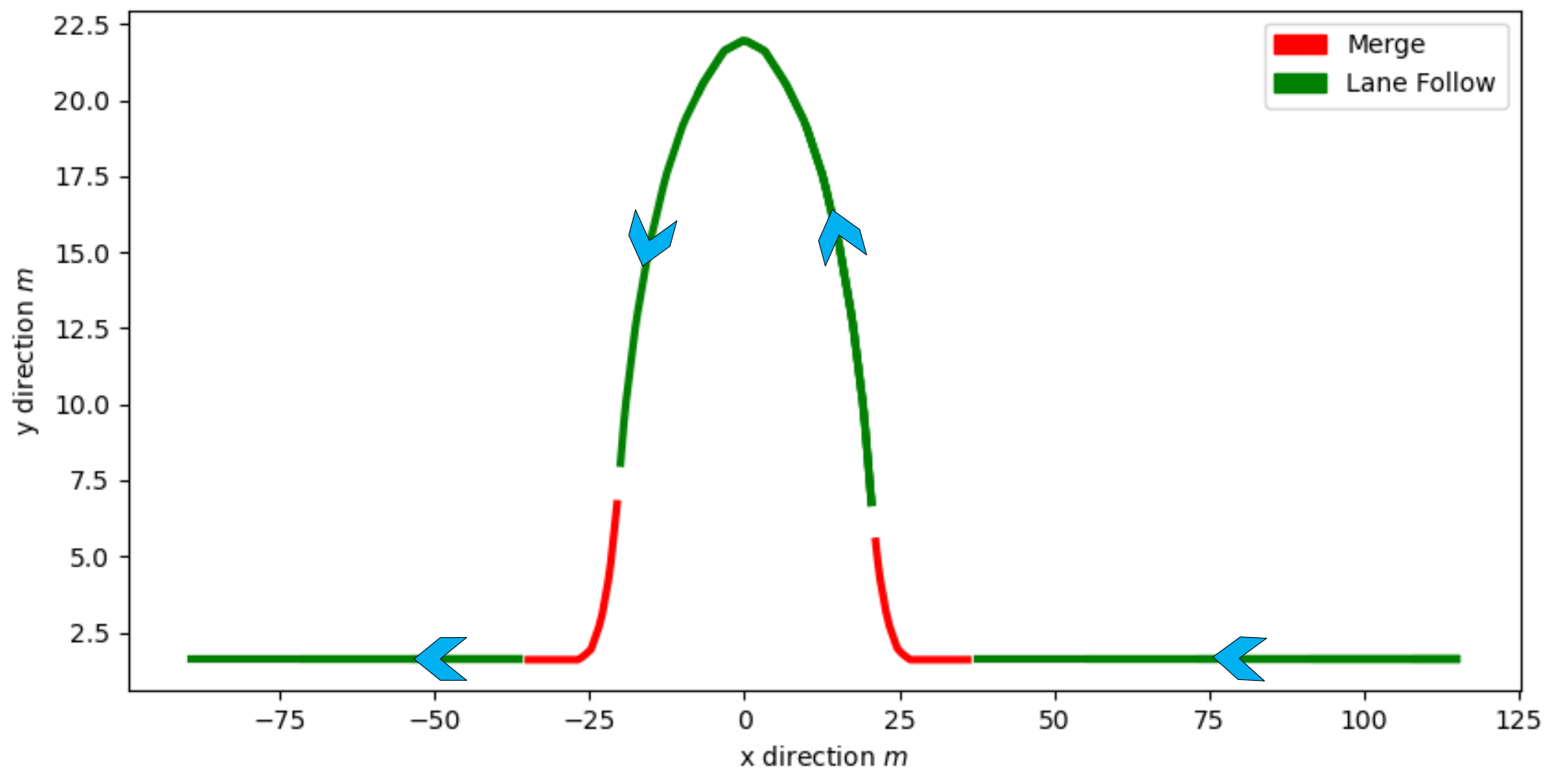

Figure 7 The Policy Prediction Result on a Vehicle Trajectory

After adding the measurement noise to the trajectory data, the policy prediction is influenced by the oscillation of the vehicle trajectory, shown in Figure 8.

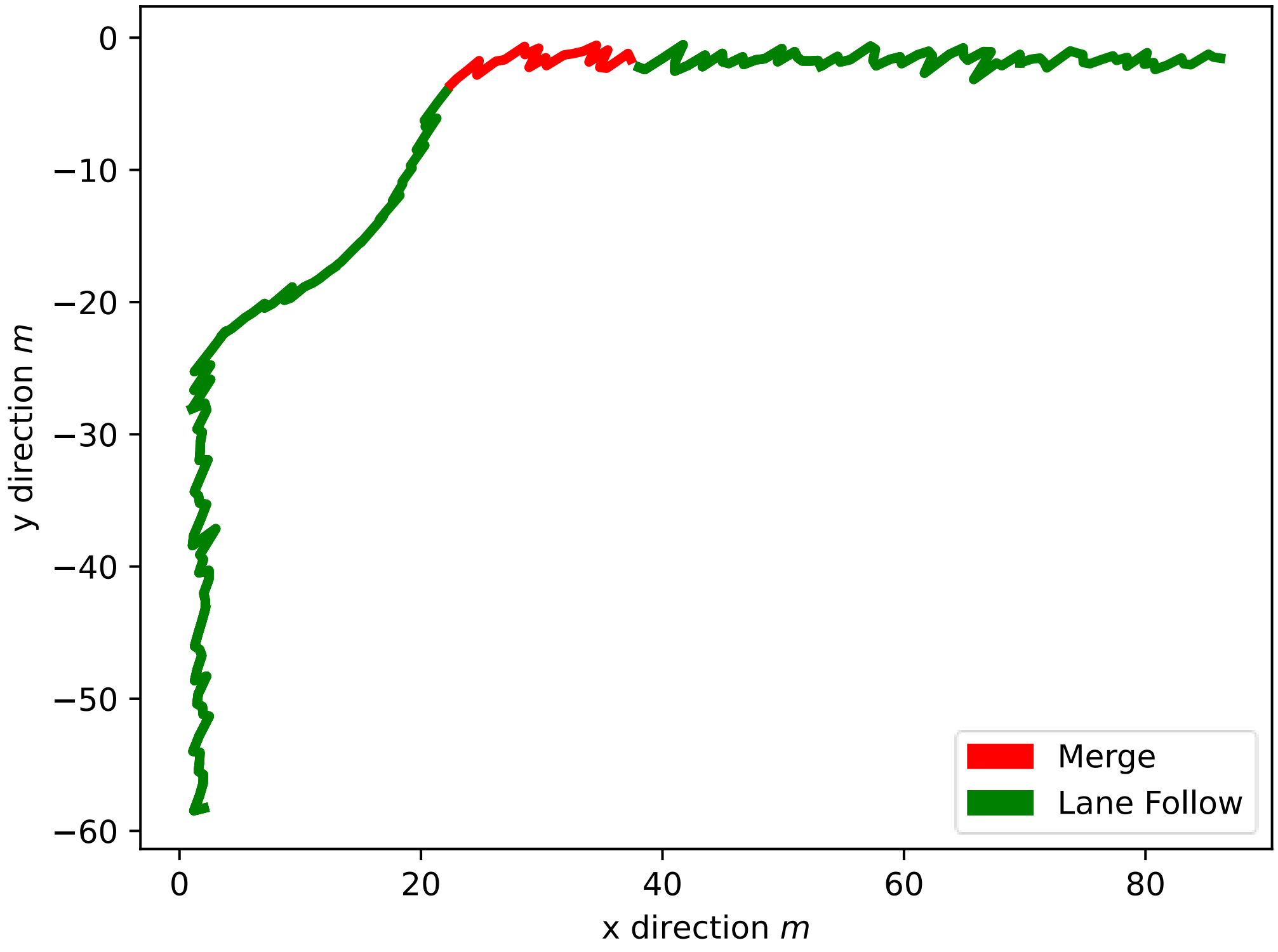

Figure 8 Policy Prediction for Noisy Vehicle Trajectory

Due to the noise in the trajectory, policy prediction cannot work properly and will cause some error and may miss the policy change between different trajectory segments are seen by comparing Figure 6 and Figure 8. Hence, we introduce the Kalman Filter for filtering out the noise along the trajectory and make sure the prediction method can work well when there is noise in the measurement in Figure 9. With the UKF, the prediction method can work better with noisy measurement of other vehicles trajectories as shown in Figure 9.

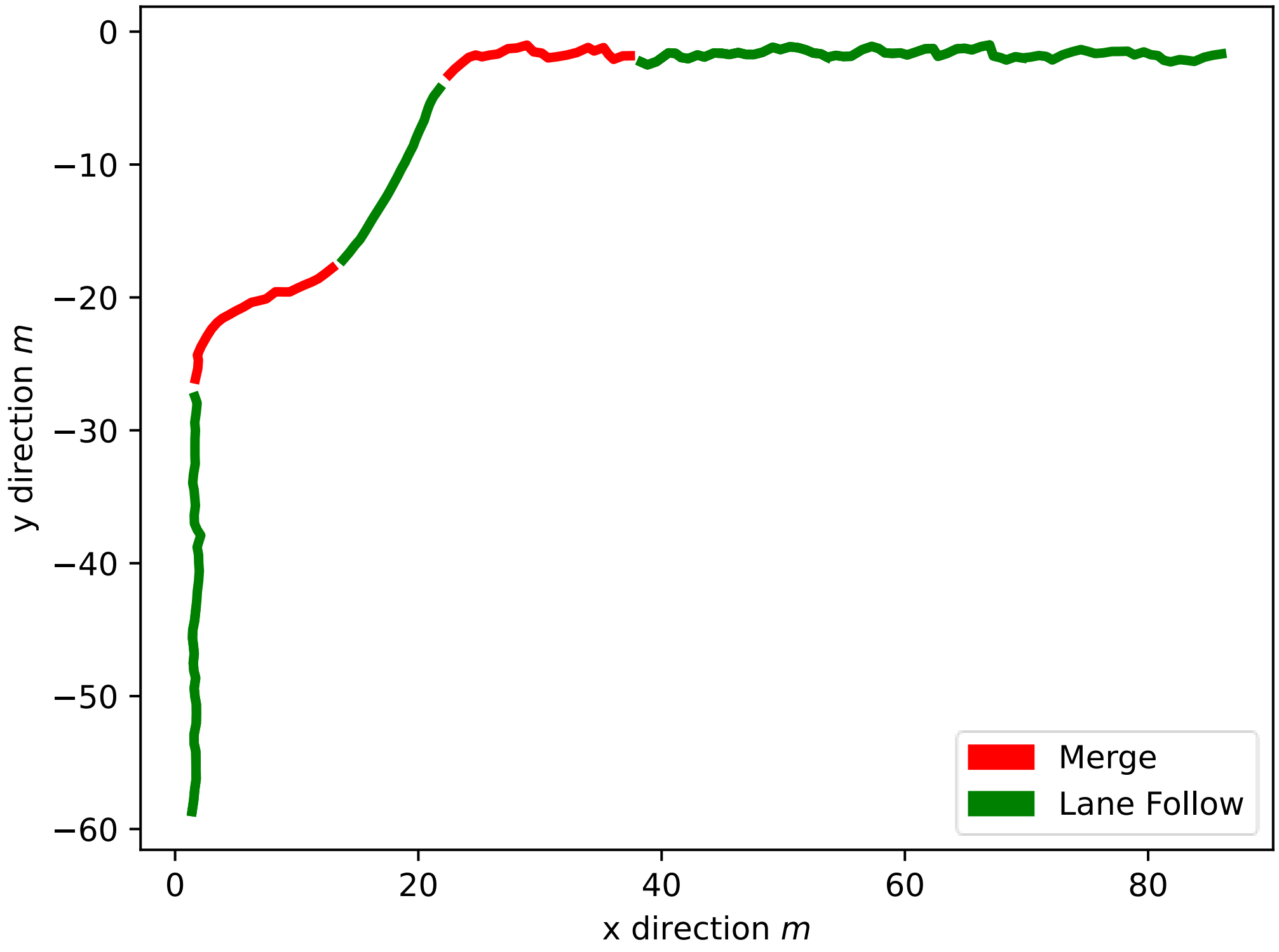

Figure 9 The Policy Prediction on Filtered Noisy Vehicle Trajectory

## 4. Vehicle Trajectory Estimation Based on UKF and Policy Prediction Method

### 4.1 Vehicle trajectory estimation

In Section 2 and Section 3, the state estimation based on the Unscented Kalman filter for unobserved internal state variables with noisy measurements and the policy prediction of the vehicle based on change point detection method are presented. Here, in this section, we introduce the method for vehicle trajectory prediction over a future period of time for a target vehicle. The whole workflow is shown in Figure 10.

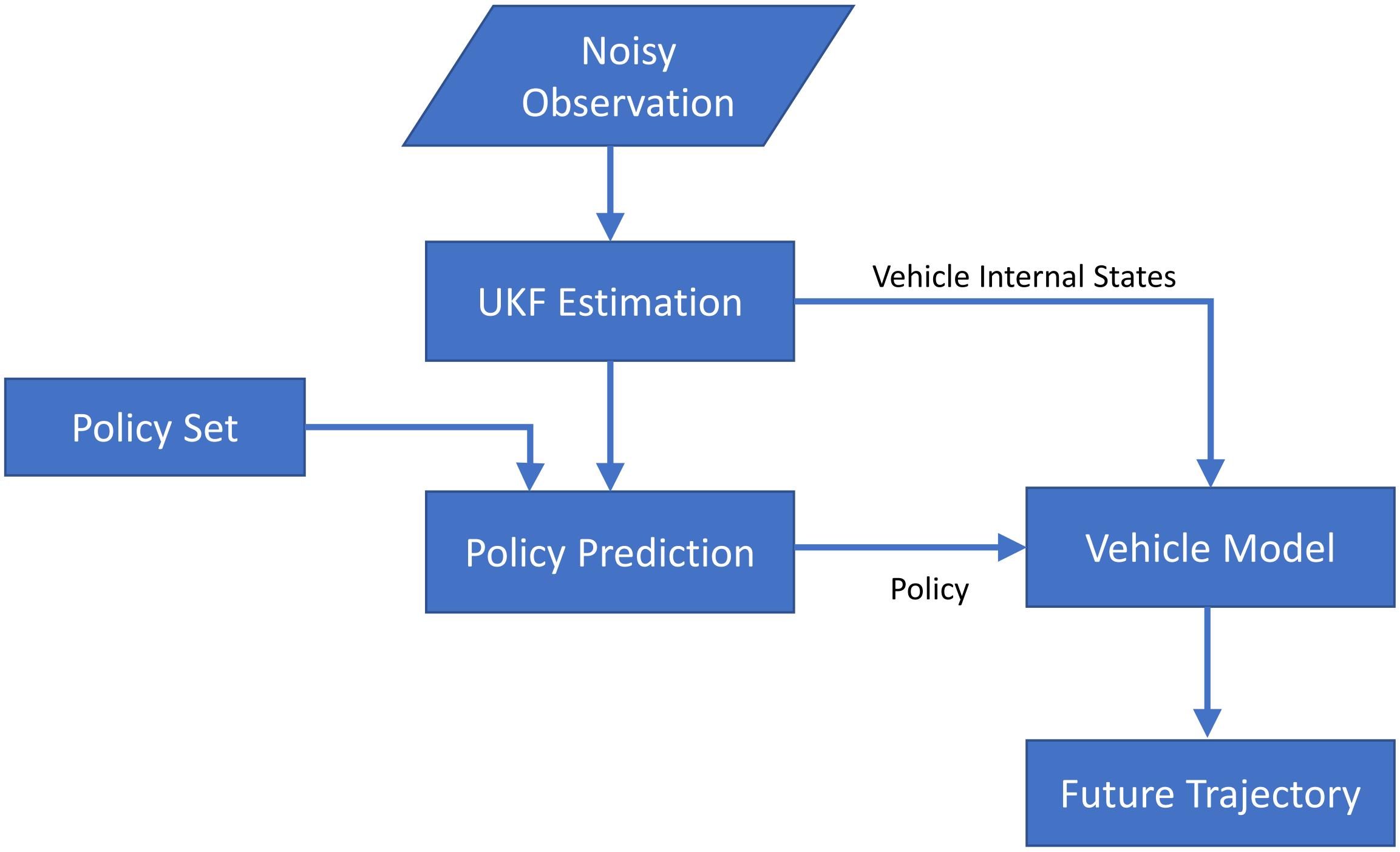

Figure10. Flowchart of the Trajectory Estimation System

The existence of the UKF estimation allows the trajectory prediction system to predict a reasonable future trajectory. Even when there is some anomaly policies, the corresponding trajectory can be retrieved. The forward simulation for determining the prediction utilized the known geometry feature of a round intersection. Given the CTRV model, the angular velocity is different at different parts of the round intersection including entering and exiting areas. The angular velocity will change during the policy execution for the policy *merge* . In contrast, the vehicles' angular velocity will remain the same for the whole stage of the policy *lane keep* . This gives us the key point of generating different simulation trajectories based on different policy.

### 4.2 Test results on the vehicle trajectory estimation method

In this section, several test results for vehicle trajectory prediction is presented.

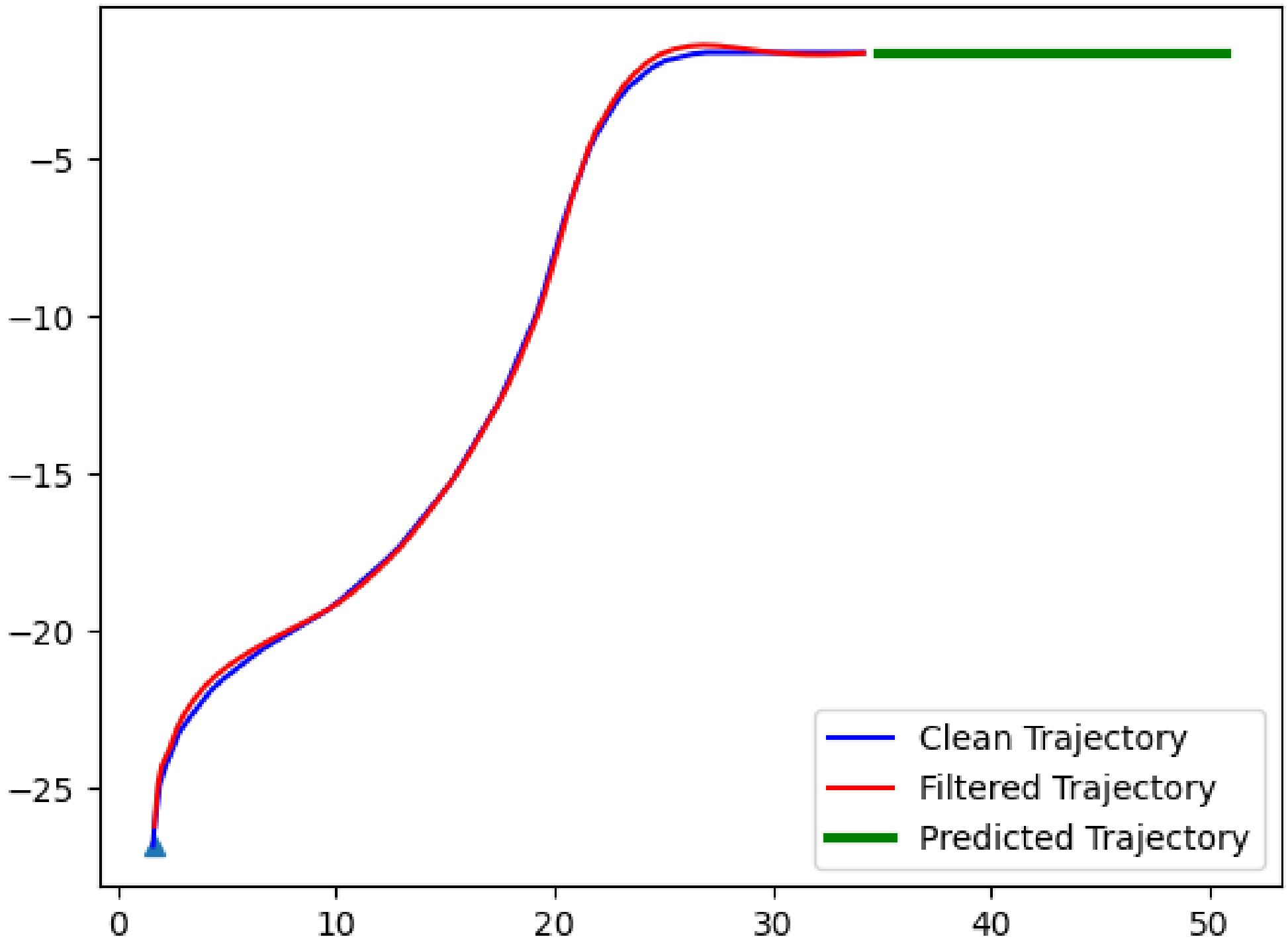

Figure 11. Trajectory Prediction Simulation Result of Vehicle Arriving at the Straight Lane Segment After the Round Intersection

In Figure 11, a vehicle trajectory estimation result is shown. The vehicle is moving from bottom to upper right of the figure and with the last state estimated with UKF and the policy prediction result as the policy is $\{lane\ \text{keep}\}$, the predicted trajectory (shown in green) is generated. With the observed trajectory, the agent vehicle doesn't have the information about angular velocity hence the UKF estimates this hidden state and provide a direction of vehicle moving forward. In Figure 12, the vehicle trajectory prediction is for the vehicle within the traffic circle, with the estimated vehicle states, the trajectory prediction method is able to forward simulate the potential vehicle path in the traffic circle.

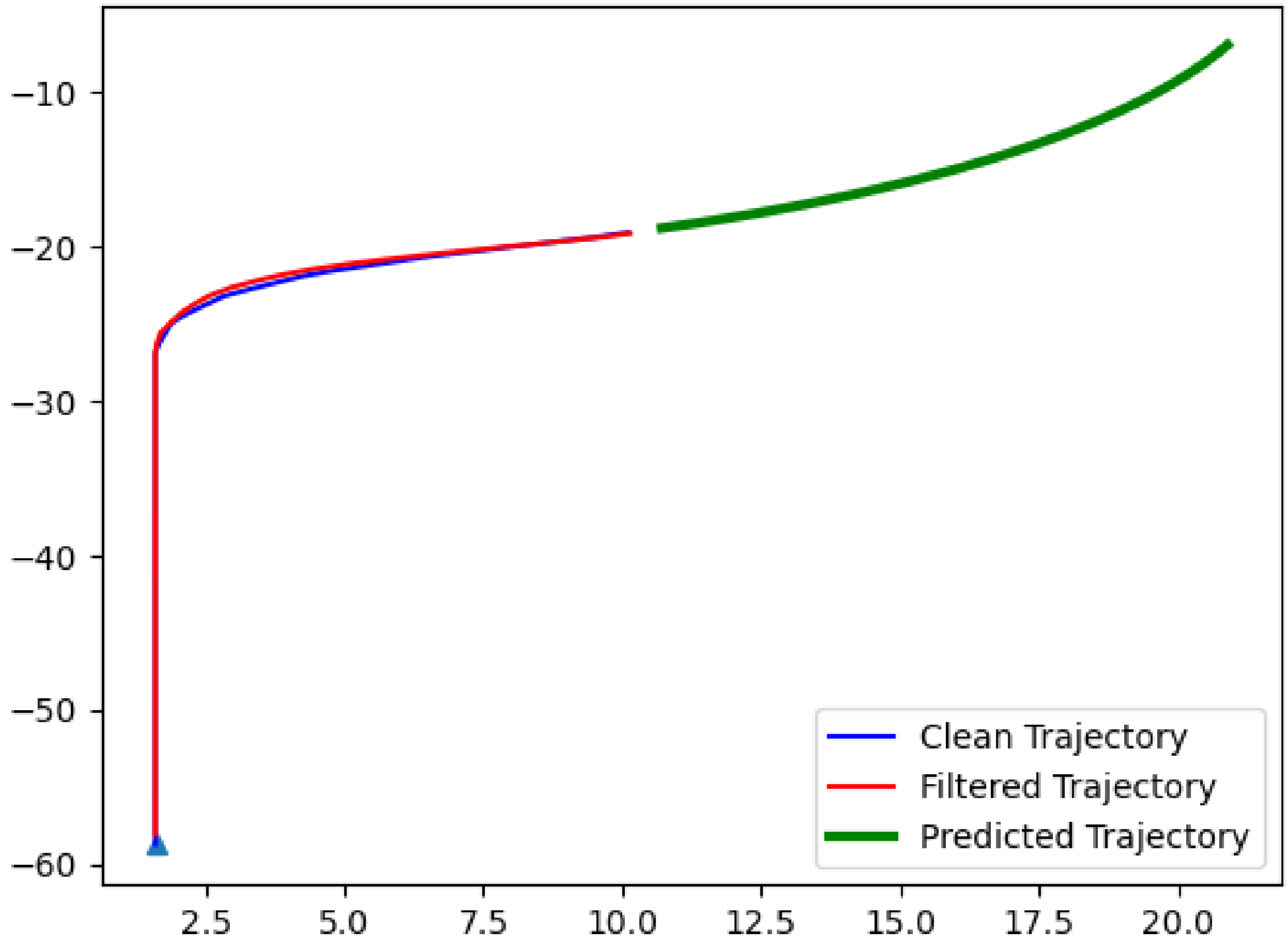

Figure 12 Trajectory Prediction Simulation Result of Vehicle Traversing the Round Intersection

**5. Conclusions**

This paper presented estimation methods on vehicle internal states, obtaining clean vehicle trajectory under noisy measurement, vehicle policy prediction for other vehicles providing a series of vehicle trajectory and proposed a method for estimating a vehicle future trajectory given a series of vehicle trajectory data.

The Unscented Kalman filter as well as the Constant Turn Rate and Velocity vehicle model provide powerful tools for estimating the vehicle trajectory given noisy measurement and also estimating the internal states of vehicles which cannot be directly captured by the sensors on the ego autonomous vehicle.

The policy prediction method based on change point detection method introduced in this chapter is a useful tool for predicting surrounding vehicle behaviors. To solve the difficult problem of estimation other vehicle's behavior, a set of latent policy shall be pre-defined so that the potential policy of other vehicles can be determined by computing the Bayesian Information Criteria (BIC) on how confident it is of the vehicle executing such policy. A good forward simulation is also necessary under different policies so that the likelihood value can be easily computed and captured for comparison. Also, computing likelihood value of Gaussian Distribution of high dimensional variables is computational heavy. An independence assumption on the trajectory component as well as the logarithm transformation makes it easier for the program to compute the likelihood value.

Vehicle trajectory estimation is an important part of vehicle decision making and planning for autonomous vehicles. Implementing the algorithms in this chapter, including Kalman filter based vehicle tracking and change point detection method based policy prediction makes the usage of series of observation data and make sure the estimation work does not require too much computation resources. And this method of estimation is straightforward, it can be easily adjusted for different traffic scenarios only if the user can generate an appropriate policy set so that the vehicle behavior can be described well enough.

Platooning or convoying of vehicles in the form of adaptive and cooperative adaptive cruise control [30], [31] on highways is a topic of high research attention. More recent work focuses on similar cooperative driving in urban roads including cooperative handling of an intersection by a convoy of cooperating vehicles [32]. While there are results for signalized intersections, corresponding results for roundabouts and traffic circles are missing. The approach in this paper can be useful for cooperative handling of roundabouts by a convoy of connected and autonomous vehicles. Active safety control systems like yaw stability controllers [33], [34] can also benefit from the vehicle state and driving intent prediction. The driving intent prediction will allow the prediction of future yaw and steering values which can be used to improve the performance of these controllers.

**Acknowledgments**

The authors thank the support of the Automated Driving Lab at Ohio State University. The authors acknowledge the partial support through the Smart Campus organization of the Ohio State University in support of the Smart Columbus project.

**References**

[1] Guvenc, L., Aksun-Guvenc, B., Emirler, M.T.: "Connected and Autonomous Vehicles," Chapter 35 in Internet of Things/Cyber-Physical Systems/Data Analytics Handbook, Editor: H. Geng, Wiley, ISBN: 978-1119173649 (2017).

[2] Guvenc, L., Aksun-Guvenc, B., Zhu, S., Gelbal, S.Y.: Autonomous Road Vehicle Path Planning and Tracking Control, Wiley / IEEE Press, Book Series on Control Systems Theory and Application, New York, ISBN: 978-1-119-74794-9 (2021).

[3] Li, X., Arul Doss, A.C., Aksun Guvenc, B., Guvenc, L.: Pre-Deployment Testing of Low Speed, Urban Road Autonomous Driving in a Simulated Environment. SAE Int. J. Adv. & Curr. Prac. in Mobility 2(6), 3301-3311 (2020). doi:10.4271/2020-01-0706.

[4] Guvenc, L., Guvenc, B.A., Demirel, B., Emirler, M.T.: Control of mechatronic systems. Institution of Engineering and Technology, (2017).

[5] Zhu, S., Gelbal, S.Y., Aksun-Guvenc, B., Guvenc, L., 2019, "Parameter-space Based Robust Gain-scheduling Design of Automated Vehicle Lateral Control," IEEE Transactions on Vehicular Technology, doi: 10.1109/TVT.2019.2937562, Vol. 68, Issue 10, pp. 9660-9671.

[6] Oncu, S., Karaman, S., Guvenc, L., Ersolmaz, S.S., Ozturk, E.S., Cetin, A.E., Sinal, M., 2007, "Robust Yaw Stability Controller Design for a Light Commercial Vehicle Using a Hardware in the Loop Steering Test Rig," IEEE Intelligent Vehicles Symposium, June 13-15, pp. 852-859.

[7] Wang, H., Tota, A., Aksun-Guvenc, B., Guvenc, L., 2018, "Real Time Implementation of Socially Acceptable Collision Avoidance of a Low Speed Autonomous Shuttle Using the Elastic Band Method," IFAC Mechatronics Journal, Volume 50, April 2018, pp. 341-355.

[8] Gelbal, S.Y., Aksun-Guvenc, B., Guvenc, L., 2020, "Elastic Band Collision Avoidance of Low Speed Autonomous Shuttles with Pedestrians," International Journal of Automotive Technology, Vol. 21, No. 4, pp. 903-917.

[9] Li, X., Zhu, S., Aksun-Güvenç, B., Güvenç, L., 2021, "Development and Evaluation of Path and Speed Profile Planning and Tracking Control for an Autonomous Shuttle Using a Realistic, Virtual Simulation Environment," Journal of Intelligent and Robotic Systems, Vol. 101, p. 42, doi.org/10.1007/s10846-021-01316-2.

[10] Aksun Guvenc, B., Guvenc, L., Karaman, S., 2009, "Robust Yaw Stability Controller Design and Hardware in the Loop Testing for a Road Vehicle," IEEE Transactions on Vehicular Technology, Vol. 58, No. 2, pp. 555-571.

[11] Ding, Y., Zhuang, W., Wang, L., Liu, J., Guvenc, L., Li, Z., 2020, "Safe and Optimal Lane Change Path Planning for Automated Driving," IMECHE Part D Passenger Vehicles, Vol. 235, No. 4, pp. 1070-1083, doi.org/10.1177/0954407020913735.

[12] Bowen, W., Gelbal, S.Y., Aksun-Guvenc, B., Guvenc, L., 2018, "Localization and Perception for Control and Decision Making of a Low Speed Autonomous Shuttle in a Campus Pilot Deployment," SAE International Journal of Connected and Automated Vehicles, doi: 10.4271/12-01-02-0003, Vol. 1, Issue 2, pp. 53-66.

[13] Meneses-Cime, K., Aksun-Guvenc, B., Guvenc, L., 2022, "Optimization of On-Demand Shared Autonomous Vehicle Deployments Utilizing Reinforcement Learning," Sensors, 22, 8317. https://doi.org/10.3390/s22218317.

[14] Bowen, W., Gelbal, S.Y., Aksun-Guvenc, B., Guvenc, L., 2018, "Localization and Perception for Control and Decision Making of a Low Speed Autonomous Shuttle in a Campus Pilot Deployment," SAE International Journal of Connected and Automated Vehicles, doi: 10.4271/12-01-02-0003, Vol. 1, Issue 2, pp. 53-66.

[15] Arul Doss, A.C., Guvenc, L., 2021, "Predicting Desired Temporal Waypoints from Camera and Route Planner Images using End-To-Mid Imitation Learning," WCX21: SAE World Congress Experience, April 13-15, Detroit, Michigan, Session AE100 ADAS and Autonomous Systems: Perception, SAE Paper Number: 2021-01-0088.

[16] Rajamani, R.: Vehicle dynamics and control. Springer Science & Business Media, (2011)

[17] Schubert, R., Richter, E., Wanielik, G.: Comparison and evaluation of advanced motion models for vehicle tracking. In: 2008 11th international conference on information fusion 2008, pp. 1-6. IEEE

[18] Bersani, M., Vignati, M., Mentasti, S., Arrigoni, S., Cheli, F.: Vehicle state estimation based on kalman filters. In: 2019 AEIT International Conference of Electrical and Electronic Technologies for Automotive (AEIT AUTOMOTIVE) 2019, pp. 1-6. IEEE

[19] Chandra, R., Guan, T., Panuganti, S., Mittal, T., Bhattacharya, U., Bera, A., Manocha, D.J.I.R., Letters, A.: Forecasting trajectory and behavior of road-agents using spectral clustering in graph-lstms. 5(3), 4882-4890 (2020).

[20] Mohamed, A., Qian, K., Elhoseiny, M., Claudel, C.: Social-stgcnn: A social spatio-temporal graph convolutional neural network for human trajectory prediction. In: Proceedings of the IEEE/CVF Conference on Computer Vision and Pattern Recognition 2020, pp. 14424-14432.

[21] Sriram, N., Liu, B., Pittaluga, F., Chandraker, M.: Smart: Simultaneous multi-agent recurrent trajectory prediction. In: European Conference on Computer Vision 2020, pp. 463-479. Springer.

[22] Wan, E.A., Van Der Merwe, R.: The unscented Kalman filter for nonlinear estimation. In: Proceedings of the IEEE 2000 Adaptive Systems for Signal Processing, Communications, and Control Symposium (Cat. No. 00EX373) 2000, pp. 153-158.

[23] Julier, S.J., Uhlmann, J.K.: New extension of the Kalman filter to nonlinear systems. In: Signal processing, sensor fusion, and target recognition VI 1997, pp. 182-193. International Society for Optics and Photonics

[24] Wan, E.A. and van der Merwe, R. (2001) The Unscented Kalman Filter. In Haykin, S., Ed., Kalman Filtering and Neural Networks, Wiley, New York, 221-280.

[25] Schubert, R., Richter, E., Wanielik, G.: Comparison and evaluation of advanced motion models for vehicle tracking. In: 2008 11th international conference on information fusion 2008, pp. 1-6. IEEE.

[26] Galceran, E., Cunningham, A.G., Eustice, R.M., Olson, E.: Multipolicy decision-making for autonomous driving via changepoint-based behavior prediction: Theory and experiment. Autonomous Robots(6), 1367 (2017). doi:10.1007/s10514-017-9619-z.

[27] Niekum, S., Osentoski, S., Atkeson, C.G., Barto, A.G.: CHAMP: Changepoint detection using approximate model parameters. In. CARNEGIE-MELLON UNIV PITTSBURGH PA ROBOTICS INST, (2014)

[28] Fearnhead, P., Liu, Z.J.S., Computing: Efficient Bayesian analysis of multiple changepoint models with dependence across segments. 21(2), 217-229 (2011).

[29] Viterbi, A.: Error bounds for convolutional codes and an asymptotically optimum decoding algorithm. IEEE transactions on Information Theory 13(2), 260-269 (1967).

[30] Emirler, M.T., Guvenc, L., Aksun-Guvenc, B., 2018, "Design and Evaluation of Robust Cooperative Adaptive Cruise Control Systems in Parameter Space," International Journal of Automotive Technology, Vol. 19, Issue 2, pp. 359-367.

[31] Ma, F., Wang, J., Zhu, S., Gelbal, S.Y., Yu, Y., Aksun-Guvenc, B., Guvenç, L., 2020, "Distributed Control of Cooperative Vehicular Platoon with Nonideal Communication Condition," IEEE Transactions on Vehicular Technology, Vol. 69, Issue 8, pp. 8207-8220, doi:10.1109/TVT.2020.2997767.

[32] Ma, F., Yang, Y., Wang, J., Li, X., Wu, G., Zhao, Y., Wu, L., Aksun-Guvenc, B., Guvenc, L., 2021, "Eco-Driving-Based Cooperative Adaptive Cruise Control of Connected Vehicles Platoon at Signalized Intersections," Transportation Research Part D: Transport and Environment, Vol. 92, 102746, ISSN 1361-9209, https://doi.org/10.1016/j.trd.2021.102746.

[33] Aksun-Guvenc, B., Guvenc, L., Ozturk, E.S., Yigit, T., 2003, "Model Regulator Based Individual Wheel Braking Control," IEEE Conference on Control Applications, İstanbul, June 23-25.

[34] Aksun-Guvenc, B., Guvenc, L., 2002, "The Limited Integrator Model Regulator and its Use in Vehicle Steering Control," Turkish Journal of Engineering and Environmental Sciences, pp. 473-482.